\documentclass[twocolumn,prl,preprintnumbers,superscriptaddress,nofootinbib]{revtex4-2}
\usepackage[utf8]{inputenc}
\usepackage{amsmath}
\usepackage{amsthm}
\usepackage{amssymb}
\usepackage{amsmath}
\usepackage{float}
\usepackage{braket}
\usepackage{graphicx}
\usepackage{url}
\usepackage{here}
\usepackage{physics}
\usepackage{here}
\usepackage{siunitx}
\usepackage{subcaption}
\usepackage{rotating}
\usepackage{comment}
\usepackage{CJKutf8}
\usepackage{bxcjkjatype}
\usepackage[colorlinks=true, pdfstartview=FitV, linkcolor=red, citecolor=blue, urlcolor=blue]{hyperref}
\usepackage{orcidlink}
\usepackage{multirow}
\usepackage[normalem]{ulem}
\usepackage{titlesec}
\usepackage{xcolor}
\usepackage{framed}
\definecolor{shadecolor}{RGB}{235,235,235}
\graphicspath{{./figures/}}

\newcommand{\be}{\begin{equation}}      
\newcommand{\ee}{\end{equation}}      
\newcommand{\bea}{\begin{eqnarray}}      
\newcommand{\eea}{\end{eqnarray}}

\newcommand{\ctext}[1]{\raise0.2ex\hbox{\textcircled{\scriptsize{#1}}}}
\usepackage{tikz}
\usetikzlibrary{quantikz}

\theoremstyle{definition}

\theoremstyle{remark}

\usepackage[normalem]{ulem}
\begin{document}
\preprint{}
\title{Timelike Quantum Energy Teleportation} 
\author{Kazuki Ikeda (\begin{uCJK}池田一毅\end{uCJK})\orcidlink{0000-0003-3821-2669}}
\email[]{kazuki.ikeda@umb.edu}
\affiliation{Department of Physics, University of Massachusetts Boston, Boston, MA 02125, USA}
\affiliation{Center for Nuclear Theory, Department of Physics and Astronomy, Stony Brook University, Stony Brook, New York 11794-3800, USA}

	\bigskip
\begin{abstract}
We establish a novel quantum protocol called Timelike Quantum Energy Teleportation (TQET) between two separated parties $A$ and $B$, designed for transporting quantum energy across spacetime. The amount of energy gained through TQET is always greater than or equal to that obtained via natural time evolution for any spin chain where $A$ and $B$ are distinguishable. This protocol uses temporal and spatial quantum correlations between agents separated by space and time. The energy supplier injects energy into the system by measuring the ground state of a many-body system that evolves over time, while the distant recipient performs a conditional operation using feedback from the supplier. When Bob acts immediately after receiving Alice's outcome, the protocol reduces to conventional QET. We present a proof-of-concept demonstration in the Ising model using quantum simulations. TQET increases energy efficiency from approximately 3\% to around 40\%, representing over a 13-fold improvement compared to QET. Furthermore, we analyzed the relationship between entanglement in time and TQET, validating the role of temporal correlations in energy activation between agents across spacetime.
\end{abstract}
\maketitle
\section{\label{sec:intro}Introduction}

Quantum energy teleportation (QET) is a protocol that leverages entanglement and local operations with classical communication (LOCC) to enable energy extraction at a remote location \cite{Hotta_2008}. 
In its simplest bipartite form, Alice (the energy supplier) performs a local measurement on the many-body ground state, thereby injecting energy into the system, and communicates her classical outcome to Bob (the energy recipient). Conditioned on this outcome, Bob applies a local unitary operation that allows him to extract energy from his subsystem. This can be viewed as a spatially separated variant of Maxwell-demon--type feedback control, implemented without any global operation on the many-body system.

QET has been experimentally demonstrated in several platforms, including high-temperature settings \cite{PhysRevLett.130.110801} and near zero temperature \cite{Ikeda:2023uni}. For a review of early theoretical developments, see \cite{Hotta:2011xj}, and for more recent progress, see \cite{Ikeda_Quantum_Energy_Teleportation_2023}. Because QET requires only local operations and classical communication, it offers an operational probe of many-body correlations and has been explored as a tool for detecting phase transitions. QET has been studied in a wide range of models, including relativistic quantum field theories \cite{2023arXiv230111712I,Ikeda:2024hbi}, spin chains with an impurity \cite{ikeda2023exploring}, and topological systems \cite{2023arXiv230209630I}. While most of these studies assume local control hardware, extensions to quantum-network settings have also been considered \cite{2023arXiv230111884I}, with applications to cryptographic tasks such as quantum zero-knowledge proof protocols \cite{Ikeda:2023yhm} and quantum key distribution \cite{Dolev:2025vjn}. Beyond entanglement, it has been suggested that quantum discord may persist after Alice's measurement \cite{Ikeda:2024hbi}, motivating complementary studies on the quantum resources underpinning QET \cite{Trevison_2015}. From an engineering perspective, QET-inspired protocols have also been discussed in the context of quantum batteries \cite{Hotta:2024jqu}. Moreover, while QET was originally formulated for energy, the underlying feedback-control idea has recently been generalized to other observables, including charge and current \cite{10.1093/ptep/ptae192}. Introducing multiple suppliers and consumers leads to a rich game-theoretic structure \cite{ikeda2025quantum}, where Nash equilibria and optimal energy-transfer mechanisms can be analyzed.

The protocol is called ``teleportation'' because, after Bob receives Alice's classical message, he can extract energy locally without waiting for energy carriers to traverse the medium.
In systems with local Hamiltonians, physical excitations and correlations propagate with a finite velocity (e.g.\ a Lieb--Robinson bound), so the temporal structure of the post-measurement dynamics is in principle relevant even when communication is idealized. Most existing studies focus on the instantaneous setting in which Bob acts immediately after receiving Alice's outcome. This naturally raises a dynamical question: can Bob extract \emph{more} energy by exploiting the time-dependent post-measurement state and its \emph{timelike correlations}, rather than restricting to the instantaneous protocol?

We refer to this dynamical extension as \emph{Timelike Quantum Energy Teleportation} (TQET). 
In TQET, Bob delays his conditional operation and applies it at a later time $t>0$, using Alice's outcome obtained at $t=0$. A priori, if Bob simply waits, his local energy will change due to the natural time evolution of the post-measurement state. The key question is whether conditional feedback applied to the dynamically evolved state can provide an \emph{additional} extraction advantage beyond what is achievable by natural time evolution alone. To isolate this genuinely feedback-enabled advantage, we compare TQET to the baseline scenario where the system is allowed to evolve after Alice's measurement but Bob performs no conditional operation. This separation is essential because, at $t>0$, the system may already carry propagating excitations generated by Alice's measurement.

QET is also interesting from the standpoint of fundamental physics---thermodynamics, many-body systems, and quantum information. However, despite more than a decade of efforts, the energy conversion efficiency (ECE) of QET is typically very low, and the amount of usable energy is severely restricted. 
Improving the extractable energy and the operational efficiency is therefore crucial for potential practical applications.

In this work, we address these two challenges. Our main accomplishments are summarized as follows:
\begin{shaded}
\begin{enumerate}
    \item We propose and demonstrate TQET as a dynamical (timelike) extension of QET.
    \item We show that TQET can yield a dramatic improvement of the operational energy conversion efficiency compared to conventional QET.
    \item We diagnose the role of timelike correlations by relating the enhanced extraction windows to time-separated correlation measures.
\end{enumerate}
\end{shaded}

\section{Timelike Quantum Energy Teleportation}
While the conventional QET is considered in the static frame, it is possible to extend the protocol to the dynamical case, where the system evolves by a natural unitary evolution $U(t)=e^{-itH}$, where $H$ is a given Hamiltonian. For simplicity, we assume that $H$ has only local interactions and local terms: $H=\sum_{n}H_n$, with $H_n$ such that $[H_n,H_m]=0$ if $|n-m|>1$. This is the common assumption in the conventional QET. 

In the conventional QET, Bob's control operation $U_B$ is applied as soon as he communicates with Alice who injects energy into the system, therefore the time-evolution of the system is not considered. However if Bob waits for a time-evolution, Bob's is affected by the system due to the relation:
\begin{equation}
    [U_{B}(b), U(t)]\neq 0,   
\end{equation}
which can be confirmed by $U(t)=\sum_{n=0}^\infty\frac{(-it)^nH^n}{n!}$ and $[U_B,H_B]\neq0$. The last condition is necessary for Bob to activate energy in the conventional QET. 

\begin{figure}
    \centering
    \includegraphics[width=\linewidth]{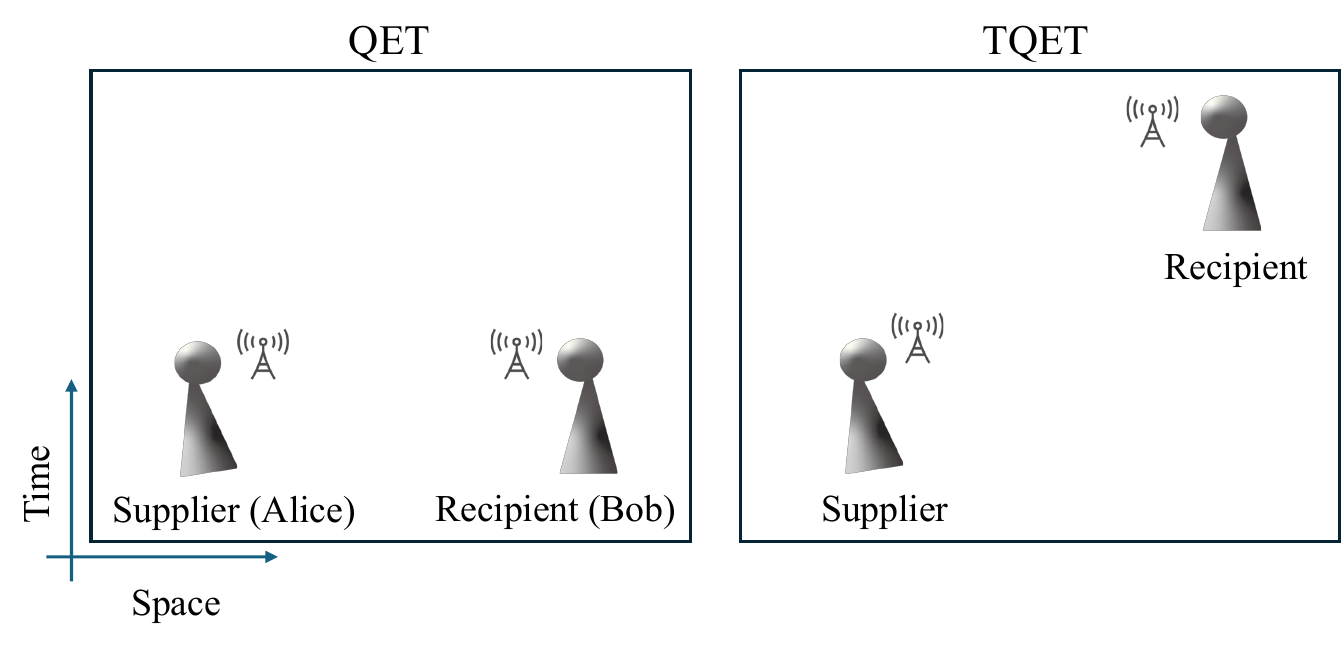}
    \includegraphics[width=\linewidth]{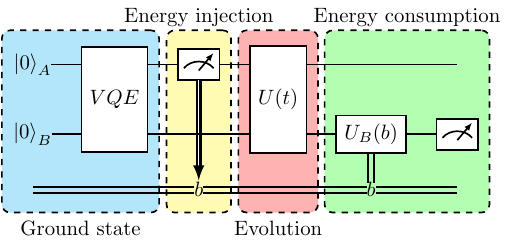}
    \captionsetup{justification=raggedright, singlelinecheck=false}
    \caption{Images of QET and TQET [top], and a TQET quantum circuit [bottom], where VQE means the variational eigensolver.}
    \label{fig:TQET_image}
\end{figure}

The protocol of TQET between two parties can be articulated as follows: 
\begin{enumerate}
    \item Alice injects energy by performing a projective measurement $P_A(b)=\frac{1+(-1)^b\sigma_A}{2}$ on the ground state at time $0$, and provides feedback $b$ to Bob.
    \item The system evolves in real time.  
    \item Using the feedback $b$ from Alice, Bob performs a conditional operation $U_B(b)$ on the state at time $t$. 
\end{enumerate}
It is important that Alice's feedback is used \textit{after} the time-evolution, which distinguishes TQET from QET. Fig.~\ref{fig:TQET_image}(top) illustrates the concept of TQET, while Fig.~\ref{fig:TQET_image} (bottom) presents a quantum circuit of TQET. 

Averaging over Alice's outcomes yields
\begin{equation}
    \rho_\text{TQET}(t)=\sum_{b\in\{0,1\}}U_B(b)U(t)P_A(b)\rho_0P_A(b)U^\dagger(t)U_B^\dagger(b). 
\end{equation}
We evaluate Bob's local energy as 
\begin{equation}
    E_\text{TQET}(t)=\Tr[(\rho_\text{TQET}(t)-\rho_0)H_B], 
\end{equation}
which is Bob's energy activated at time $t$. The conventional QET is defined at $t=0$, where his local energy decreases in comparison to the ground state energy: $E_\text{QET}:=E_\text{TQET}(0)<0$. It is important that $E_\text{TQET}(t)$ can be either negative or positive, depending on $t$. When it is negative, it indicates energy teleportation. 

We consider the post-measurement state
\begin{equation}
    \rho_A=\sum_{b\in\{0,1\}}P_A(b)\rho_0P_A(b) 
\end{equation}
and its natural time-evolution (NTE) driven by $H$:
\begin{equation}
    \rho_A(t)=U(t)\rho_AU^\dagger(t). 
\end{equation}
This describes Bob's local energy expectation value, using the time-evolved state after Alice's measurement:
\begin{equation}
    E_\text{NTE}(t)=\Tr[(\rho_A(t)-\rho_0)H_B].
\end{equation}
Since $[P_A,H_B]=0$, $E_\text{NTE}$ is strictly 0 at $t=0$.

We quantify the \emph{net} contribution of the classical feedback beyond ordinary transport by
\begin{align}
\begin{aligned}
\label{eq:TQET_NTE_diff}
\Delta E(t,\theta)
&:=E_\text{TQET}(t,\theta)-E_\text{NTE}(t)\\
&=\sum_{b\in\{0,1\}} \Tr\!\left[\rho_b(t)\big(U_B^\dagger(b)H_B U_B(b)-H_B\big)\right],
\end{aligned}
\end{align}
where $\rho_b(t)=U(t)P_A(b)\rho_0 P_A(b)U^\dagger(t)$ are the (unnormalized) branch states.
We will mainly consider the \emph{optimized} value $\Delta E_{\min}(t):=\min_\theta \Delta E(t,\theta)$. Since $\theta=0$ (i.e., doing nothing) gives $\Delta E(t,0)=0$, we always have $\Delta E_{\min}(t)\le 0$. At $t=0$, $E_\text{NTE}(0)=0$ and therefore $\Delta E_{\min}(0)=E_\text{QET}$.

Basic properties of TQET are summarized as follows:
\begin{shaded}
\begin{itemize}
    \item Both $E_\text{NTE}(t)$ and $E_\text{TQET}(t,\theta)$ may take either sign depending on $t$ and $\theta$.
    \item For the optimized protocol, \[\Delta E_{\min}(t)=\min_\theta\{E_\text{TQET}(t,\theta)-E_\text{NTE}(t)\}\le 0\] for any $t$, with strict inequality whenever $N(t)\neq 0$.
    \item If $\min_t E_\text{TQET}(t,\theta^\star(t))<E_\text{QET}$, the time-delayed protocol yields strictly larger extractable energy than the instantaneous QET protocol.
    \item For $t>0$, $E_\text{TQET}$ generally contains both (i) the baseline energy redistribution caused by Alice's measurement and subsequent unitary dynamics and (ii) the additional extraction enabled by conditional feedback. Our $\Delta E(t)$ isolates contribution (ii).
\end{itemize}
\end{shaded}

\subsection*{Analytical formula of TQET and optimal control}
We now derive a \emph{model–independent} expression for the energy balance of TQET that holds for any spin-chain model.

For TQET, Bob applies the conditional unitary $U_B(b)=\exp\!\big(-i(-1)^b\theta\sigma_B\big)$ using Alice's classical bit $b\in\{0,1\}$. A direct computation using $\sigma_B^2=I$ gives the analytical identity~\eqref{eq:TQET_NTE_diff}:
\begin{equation}
\label{eq:exact-dE}
\Delta E(t,\theta)\;=\;\frac{1}{2}\big[\cos(2\theta)-1\big]\,M(t)\;+\;\frac{1}{2}\sin(2\theta)\,N(t),
\end{equation}
with
\begin{align}
\begin{aligned}
\label{eq:M-def}
M(t) &:= \frac{1}{2}\Tr\Big(\rho_A(t)[\sigma_B,[\sigma_B, H_B]]\Big),\\
N(t) &:= \frac{i}{2}\Tr\Big(U(t)\{\sigma_A,\rho_0\}U^\dagger(t)[\sigma_B,H_B]\Big).
\end{aligned}
\end{align}

Minimizing \eqref{eq:exact-dE} with respect to~$\theta$ yields
\begin{align}
\begin{aligned}
\label{eq:parameter}
\cos\!\big(2\theta^\star(t)\big)&=-\frac{M(t)}{\sqrt{M(t)^2+N(t)^2}},\\
\sin\!\big(2\theta^\star(t)\big)&=-\frac{N(t)}{\sqrt{M(t)^2+N(t)^2}}
\end{aligned}
\end{align}
and the optimized analytical TQET-energy formula for Bob:
\begin{equation}
\label{eq:theta-star}
    \boxed{\Delta E_{\min}(t)=\frac{-M(t)-\sqrt{M(t)^2+N(t)^2}}{2}\le 0.}
\end{equation}
In conclusion, $\Delta E_{\min}(t)$ is \emph{negative} for any $t$ such that $N(t)\neq 0$, which means that Bob can extract more energy than what would be possible through natural time evolution alone. At $t=0$, this formula reduces to the conventional QET energy.

\begin{figure*}
\centering
\setkeys{Gin}{width=\linewidth}
\begin{subfigure}{0.32\textwidth}
\includegraphics{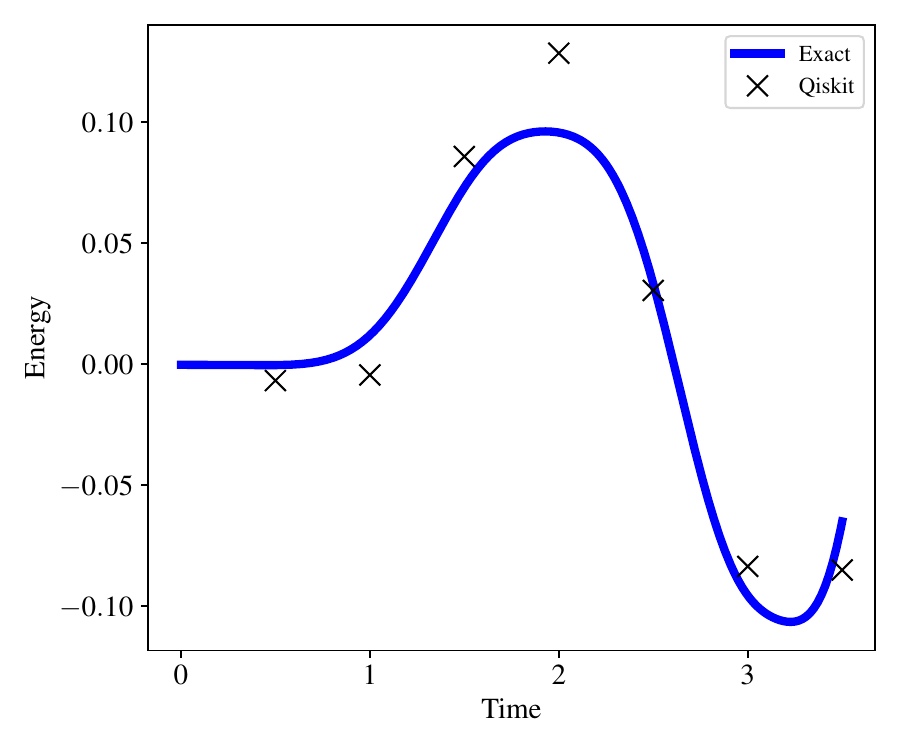}
\caption{$E_\text{TQET}(t)$}
\label{fig:cdiagram}
\end{subfigure}
\hfil
\begin{subfigure}{0.32\textwidth}
\includegraphics{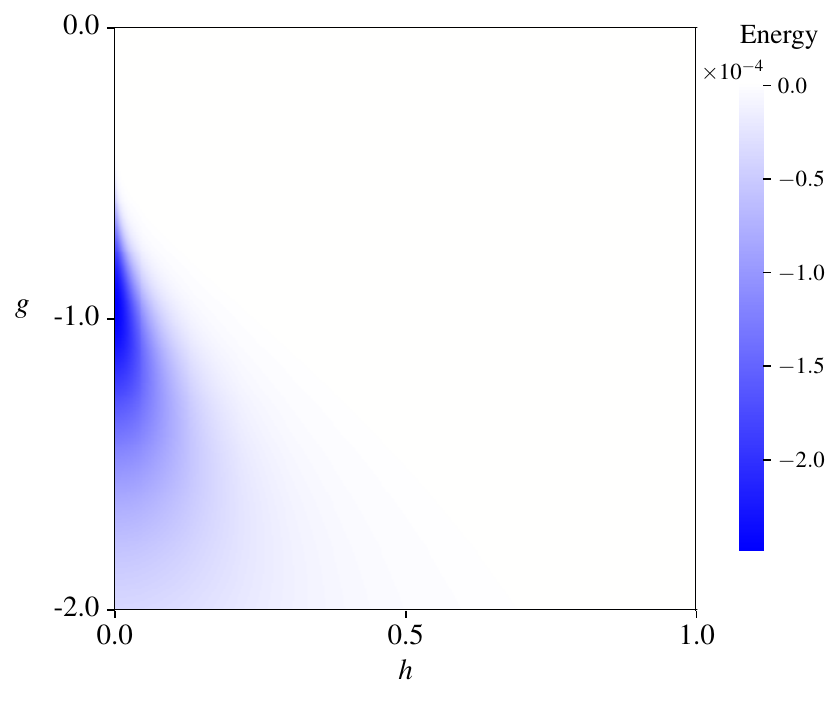}
\caption{QET energy}
\label{fig:capparatus}
\end{subfigure}%
\hfil
\begin{subfigure}{0.32\textwidth}
\includegraphics{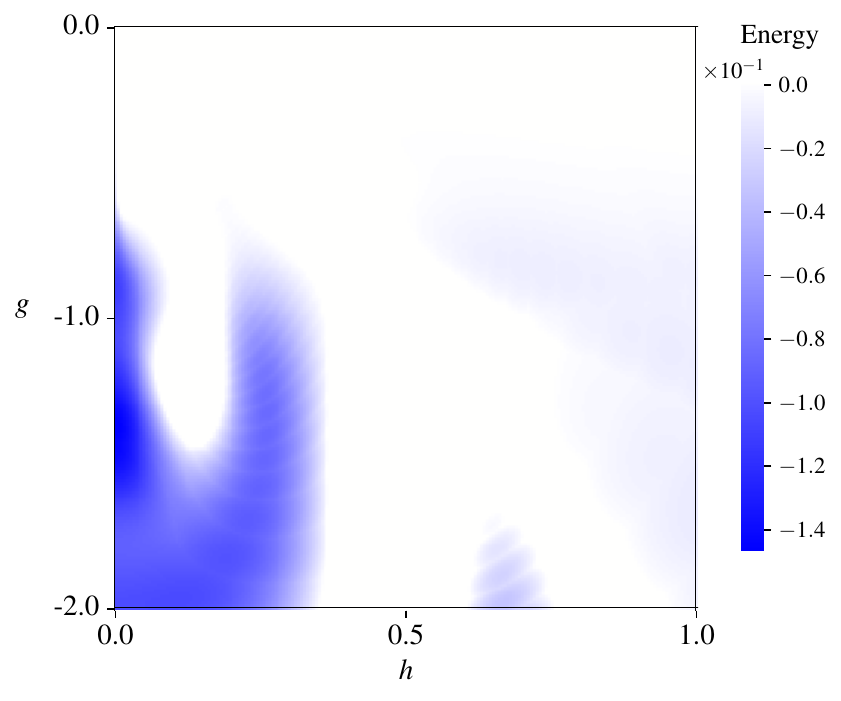}
\caption{TQET energy} 
\label{fig:cdiagram}
\end{subfigure}%
\hfil
\begin{subfigure}{0.49\textwidth}
\includegraphics{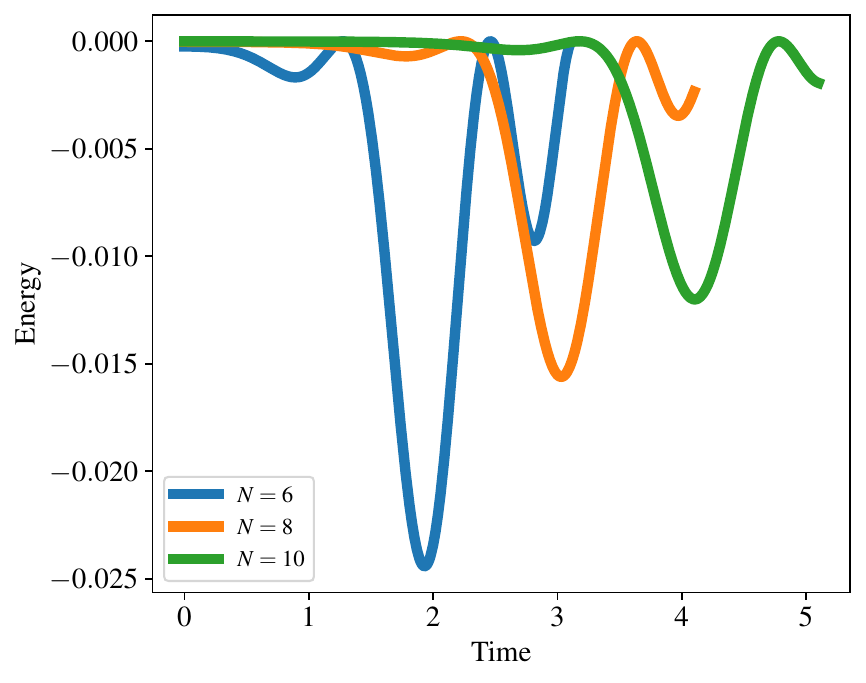}
\caption{$\Delta E(t)$}
\label{fig:cdiagram}
\end{subfigure}
\hfil
\begin{subfigure}{0.49\textwidth}
\includegraphics{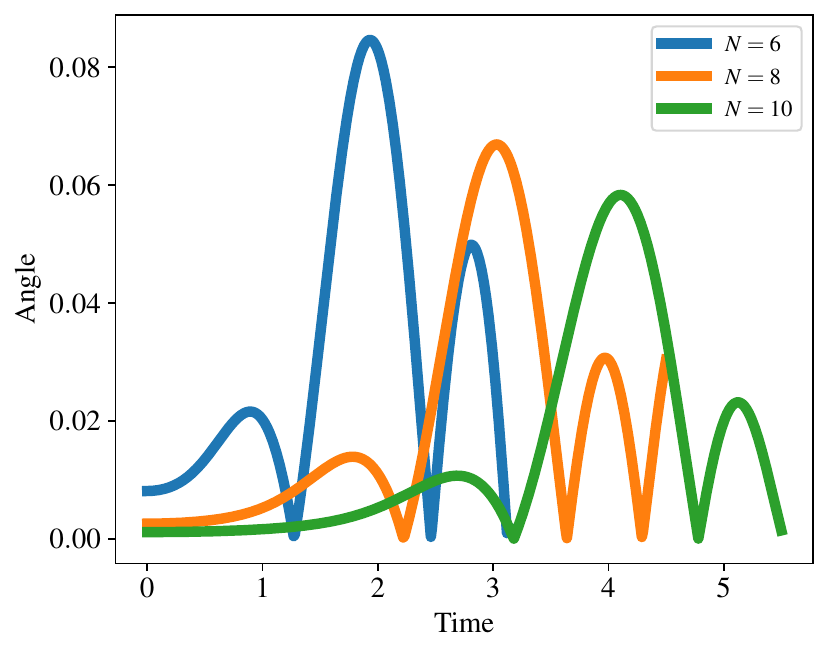}
\caption{Optimized angle $\theta^*$}
\label{fig:cdiagram}
\end{subfigure}
\caption{TQET and QET energies teleported to Bob at $n_B=N-1$ from Alice at $n_A=2$ within the system of $N=6$.}
\label{fig:DQET_energy}
\end{figure*}

\section{Model}
We consider the following Ising model with transverse and longitudinal fields:
\begin{equation}
    H=-J\sum_{n=1}^{N-1}Z_nZ_{n+1}-h\sum_{n=1}^NZ_n-g\sum_{n=1}^NX_n. 
\end{equation}
Bob's Hamiltonian is defined as
\begin{equation}
    H_B=-JZ_{B}(Z_{B-1}+Z_{B+1})-hZ_B-gX_B.
\end{equation}
When $h=0,g=0$, it is the classical Ising model and when $h=0,g=1$, it is the transverse Ising model. Both are integrable systems. The model becomes chaotic if $h/J=0.5,g/J=-1.05$. Unless specified, we work with $h/J=0,g/J=-1.05$ throughout the work. 

We use $\sigma_A=Z_A$ for Alice's measurement $P_A(b)=\frac{1+(-1)^b\sigma_A}{2}$ and $\sigma_B=Y_B$ for Bob's conditional operation $U_B(b)=e^{-i(-1)^b\theta\sigma_B}$, where $\theta\in\mathbb{R}\setminus\{0\}$ is chosen so that TQET becomes optimal. Clearly, Bob's operation does not commute with the Hamiltonian $[\sigma_B,H]\neq0$. Therefore the amount of energy teleported to Bob depends on his timing to apply $U_B(b)$ to the system. 

To ensure non-triviality of the protocol, it is important that Alice's projective measurement $P_A$ does not directly affect Bob's local energy, since $[P_A,H_B]=0$. Moreover, it is only $X_A$ that is non-commutative with $P_A$. Therefore, Alice's Hamiltonian $H_A$ can be defined as $H_A = -g X_A$.

Fig.\ref{fig:DQET_energy} (a) demonstrates TQET, using a quantum simulator from \texttt{Qiskit} \cite{Qiskit}. We show heatmaps of the teleported energy by (b) QET and (c) TQET, where the minimized energy over time $\min_tE_\text{TQET}(t)<0$ is plotted. For all $g,h$, $\min_tE_\text{TQET}(t)\le E_\text{QET}$ holds true, and at certain points, $E_\text{TQET}$ is considerably smaller than $E_\text{QET}$, suggesting that TQET can teleport approximately 1,000 times more energy than QET.

Panel (d) plots $\Delta E(t)$, which is always non-positive. Local minima of $\Delta E(t)$ correspond to times when correlations from $A$ have reached $B$ and the drive $i[\sigma_B,H]$ is phase-aligned with the incoming excitation; this is when $|N(t)|/\sqrt{M(t)^2+N(t)^2}$ is largest, which can be confirmed in panel (e), where the optimized $\theta^*(t)$ is plotted. Because $M(t)$ is set by Bob's locality while $N(t)$ grows once correlations propagate, negative $\Delta E(t)$ is generic past a light-cone delay, and small timing errors only perturb quadratically around each optimum.

\section{Energy Conversion Efficiency}
In practice, only the \emph{net} contribution of conditional feedback beyond natural time evolution should be credited. We therefore define the operational energy conversion efficiency (ECE) as
\begin{equation}
\label{eq:ECE}
\widehat{\eta}\;:=\;\frac{\displaystyle \max_{t}\Big[-\,\Delta E_{\min}(t)\Big]}{E_{\mathrm{input}}},
\end{equation}
with $E_{\mathrm{input}}=\Tr\!\big[(\rho_A-\rho_0)\,H_A\big]$ and $\Delta E_{\min}(t)$ from \eqref{eq:theta-star}. This choice subtracts NTE and matches the efficiency reported in our figures. Eq.~\eqref{eq:theta-star} guarantees $\Delta E_{\min}(t)< 0$ for any $t$ with $N(t)\neq 0$. Unoptimized $\Delta E(t,\theta)$ may take either sign; in plots we report the minimized value using the optimized parameters \eqref{eq:parameter}.

\begin{figure}[H]
    \centering
    \includegraphics[width=\linewidth]{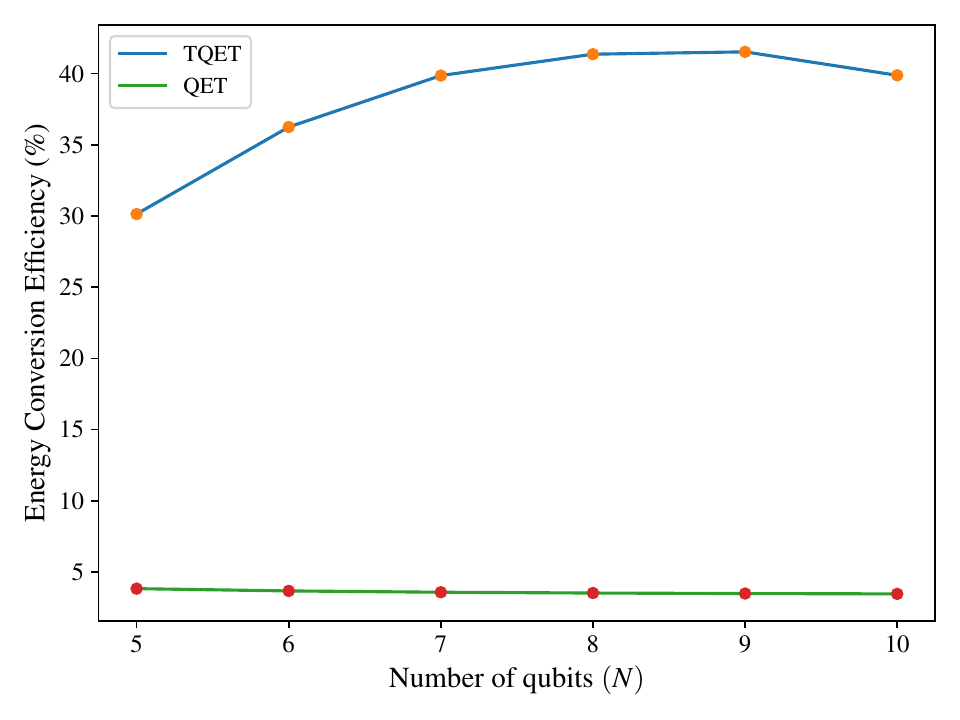}
    \caption{The $N$-dependence of the energy conversion efficiency \eqref{eq:ECE} of TQET, compared to QET.}
    \label{fig:Efficiency}
\end{figure}

In our spin-chain system, the optimal energy distribution can be done by putting Alice on $n=2$~\cite{ikeda2025quantum}. In Fig.~\ref{fig:Efficiency}, we display the ECE of TQET at the moment when the teleported energy reaches its minimum level at each site. The operational efficiency credits only the \emph{net} extraction, with $\widehat{\eta}=\max_t[-\Delta E_{\min}(t)]/E_{\mathrm{input}}$. The $\sim40\%$ value for TQET compared to $\sim3\%$ for QET indicates that, at the optimal windows, the predictor $N(t)$ is large enough to overcome the local cost $M(t)$ by an order of magnitude. The saturation of $\widehat{\eta}$ with $N$ reflects that $E_{\mathrm{input}}$ (set near $A$) and the optimized local extraction near $B$ are both bounded by local operator norms; the advantage arises from phase-matched timing, not from indefinitely accumulating energy.

\begin{figure}[H]
    \centering
    \includegraphics[width=\linewidth]{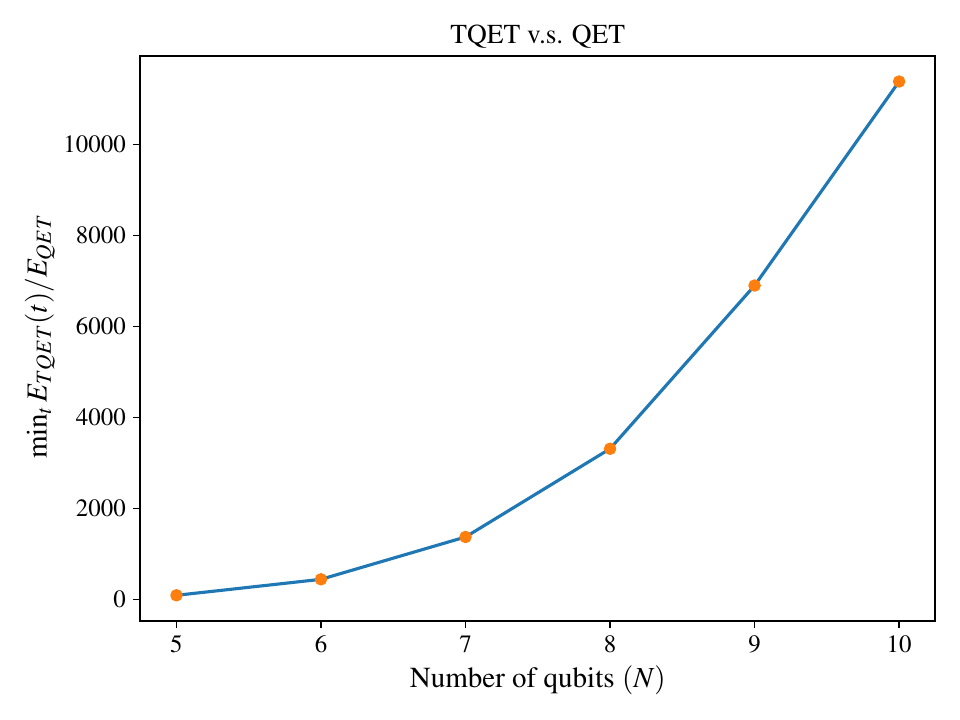}
    \caption{A comparison of the net energies teleported by TQET and QET was evaluated using eq.~\eqref{eq:TQET_ratio}. }
    \label{fig:TQET_vs_QET}
\end{figure}

To explore the scalability of TQET, here we assign Bob to $n_B=N-1$ $(|n_A-n_B|=N-3)$ in an $N$-qubit Ising model. In Fig.~\ref{fig:TQET_vs_QET}, we compare Bob's local energy by TQET, with the results from QET. This comparison is evaluated by the following form:
\begin{equation}
\label{eq:TQET_ratio}
    \frac{\min_tE_\text{TQET}(t)}{E_\text{QET}}.
\end{equation}
It is crucial that as the system size increases, the energy extractable by TQET significantly surpasses that of QET. Notably, this implies that TQET is more effective at longer distances, addressing the limitation of QET, where the extractable energy diminishes as the distance from Alice increases. Once $N(t)$ develops at $B$, the conditional drive recovers much of the lost performance. This behavior is consistent with a Lieb--Robinson picture: extraction windows open once the retarded commutator (entering $N(t)$) becomes appreciable at $B$.

\begin{figure}[H]
    \centering
    \includegraphics[width=\linewidth]{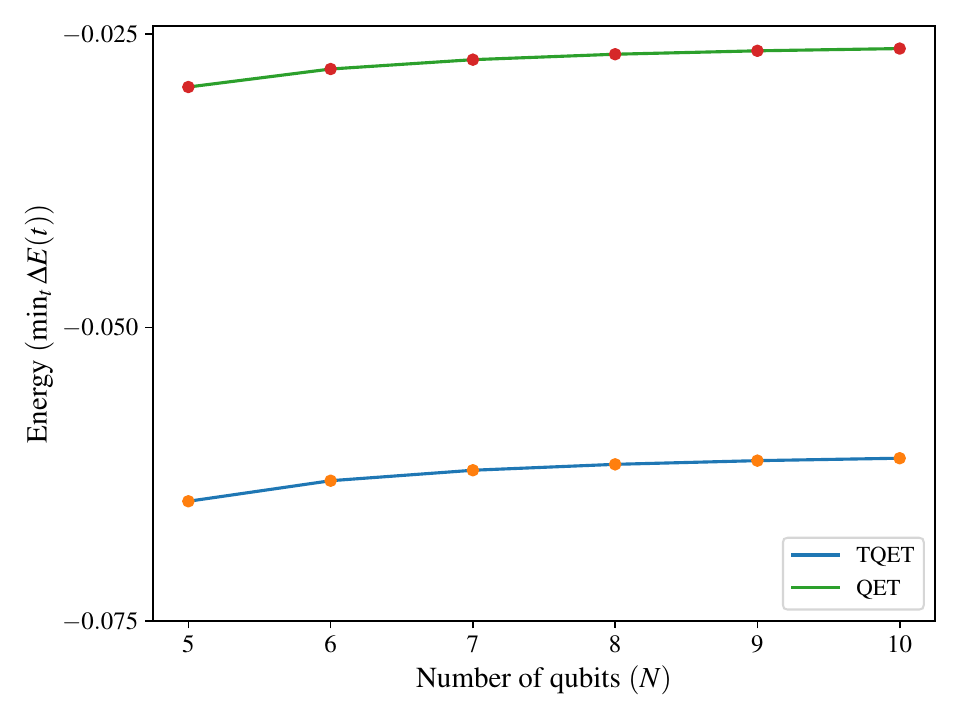}
    \caption{A comparison between QET and TQET.}
    \label{fig:Energy_TQET_QET}
\end{figure}

Furthermore, we evaluate the maximal energy induced purely by TQET. We compare the energy of QET with the quantity defined in the following manner (labeled as ``TQET" in the figure legend):
\begin{equation}
\label{eq:TQET_min}
    \min_t \Delta E(t)~\text{for}~E_\text{TQET}(t)<0, 
\end{equation}
where $\Delta E(t)$ is defined by eq.~\eqref{eq:TQET_NTE_diff}. Fig.~\ref{fig:Energy_TQET_QET} depicts the comparison between TQET and QET when Bob is positioned next to Alice, enabling an analysis of how teleported energy depends on system size while maintaining a constant distance between Alice and Bob. Notably, the QET energy is always surpassed by the minimal energy extracted via TQET (as defined in Eq.~\eqref{eq:TQET_min}), showing that TQET consistently provides a greater energy extraction advantage across all system sizes, with the energy gap between TQET and QET remaining constant regardless of $N$.

\begin{figure*}
\centering
\setkeys{Gin}{width=\linewidth}
\begin{subfigure}{0.32\textwidth}
\includegraphics[width=\linewidth]{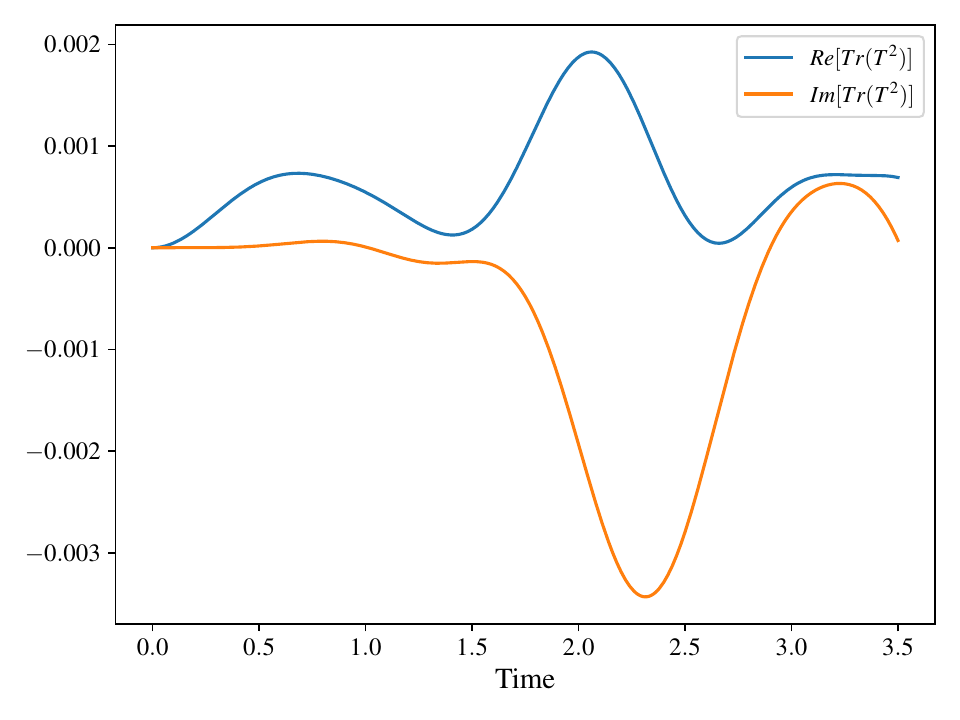}
\caption{$\Tr T^2(t,\rho_A)$}
\label{fig:capparatus}
\end{subfigure}%
\hfil
\begin{subfigure}{0.32\textwidth}
\includegraphics[width=\linewidth]{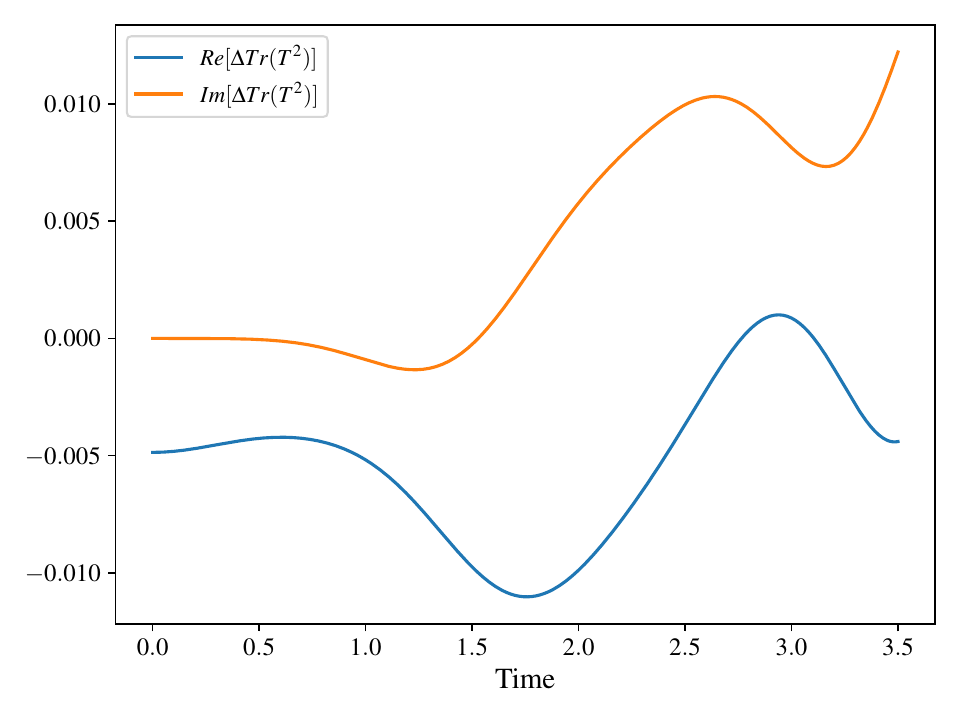}
\caption{$\Delta \Tr T^2(t)$} 
\label{fig:cdiagram}
\end{subfigure}%
\hfil
\begin{subfigure}{0.32\textwidth}
\includegraphics[width=\linewidth]{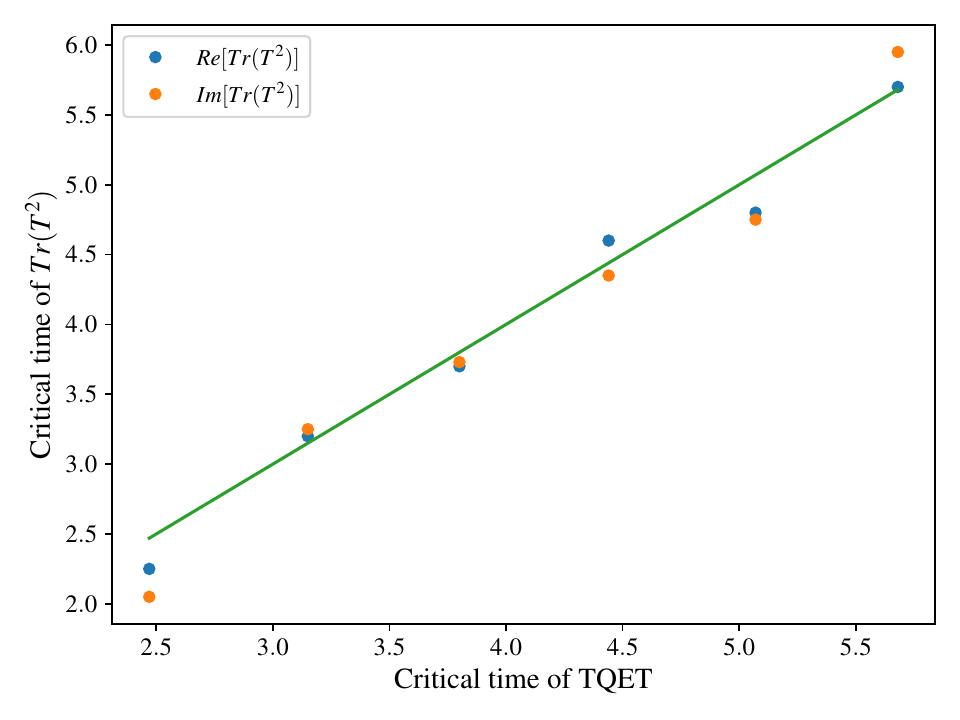}
\caption{TQET vs. $\Delta\Tr[T^2]$} 
\label{fig:cdiagram}
\end{subfigure}%
\label{fig:cdiagram}
\caption{The entanglement entropy in time between Alice and Bob.}
\label{fig:Timelike_EE}
\end{figure*}

\section{Time-Separated Correlations}

We now relate the time-dependent advantage of TQET to the build-up of correlations between Alice's region $A$ and Bob's region $B$ at different times. Consider Heisenberg operators supported on $A$ and $B$,
\begin{equation}
    O_A(t)=U^\dagger(t)\,O_A\,U(t),\qquad O_B(0)=O_B,
\end{equation}
and the time-separated correlation functions
\begin{equation}
    C_{O_A,O_B}(t;\rho):=\Tr\!\big[\rho\,O_A(t)\,O_B(0)\big].
\end{equation}

Following the timelike (spacetime) density-matrix formalism of \cite{Milekhin:2025ycm}, we introduce the spacetime density matrix $T_{AB}(t;\rho)$ associated with the initial state $\rho$ and the unitary evolution $U(t)$. It is defined by the property that for any operators $O_A$ on $A$ and $O_B$ on $B$,
\begin{equation}
\label{eq:STDM_def}
    \Tr\!\big[\rho\,O_A(t)\,O_B(0)\big]
    =\Tr\!\Big[T_{AB}(t;\rho)\,\big(O_A\otimes O_B\big)\Big].
\end{equation}
In contrast to an ordinary reduced density matrix, $T_{AB}(t;\rho)$ is generally \emph{not} Hermitian, reflecting the fact that operator ordering becomes nontrivial for timelike separations \cite{Milekhin:2025ycm}. Consequently, its moments such as $\Tr[T_{AB}^n]$ can be complex.
In this work we focus on the second moment $\Tr[T_{AB}^2]$, which provides a compact diagnostic of timelike correlations and can be measured by feasible protocols \cite{Milekhin:2025ycm}.

For numerical evaluation, we choose complete orthonormal sets of Hermitian operators $\{O_{A,\alpha}\}$ on $A$ and $\{O_{B,\beta}\}$ on $B$, normalized by the Hilbert--Schmidt inner product
\begin{equation}
    \Tr(O_{X,\mu}O_{X,\nu})=\delta_{\mu\nu}\qquad (X=A,B).
\end{equation}
(For qubit subsystems one may take properly normalized Pauli-string bases.)
Define the correlator matrix
\begin{equation}
\label{eq:C_alphabeta_def}
    C_{\alpha\beta}(t;\rho):=\Tr\!\big[\rho\,O_{A,\alpha}(t)\,O_{B,\beta}(0)\big].
\end{equation}
Then Ref.~\cite{Milekhin:2025ycm} shows (see Appendix~H therein) that the second moment admits the correlator expansion
\begin{equation}
\label{eq:TrT2_corr}
    \Tr\!\big[T_{AB}^2(t;\rho)\big]=\sum_{\alpha,\beta} \big(C_{\alpha\beta}(t;\rho)\big)^2,
\end{equation}
which is generically complex, while the nonnegative quantity
\begin{equation}
\label{eq:TrTTdag_corr}
    \Tr\!\big[T_{AB}(t;\rho)\,T_{AB}^\dagger(t;\rho)\big]
    =\sum_{\alpha,\beta}\big|C_{\alpha\beta}(t;\rho)\big|^2
\end{equation}
measures the Hilbert--Schmidt norm of the time-separated correlator matrix.

We evaluate $\Tr[T_{AB}^2(t;\rho)]$ for two initial states: the ground state $\rho_0$ and the post-measurement state $\rho_A=\sum_{b}P_A(b)\rho_0 P_A(b)$. In the following analysis we take parameters $N=6$ and $|n_A-n_B|=3$. Figure~\ref{fig:Timelike_EE}(a) shows the time evolution of $\Tr[T_{AB}^2(t;\rho_A)]$. Consistent with the general properties of $T_{AB}$, the second moment develops a nontrivial complex phase: its real part and imaginary part vary in time and do not vanish identically. This indicates that time-separated correlations between $A$ and $B$ are present throughout the evolution.

To isolate the effect induced by Alice's measurement, we consider
\begin{equation}
\label{eq:DeltaTrT2_def}
    \Delta \Tr T^2(t):=\Tr\!\big[T_{AB}^2(t;\rho_A)\big]-\Tr\!\big[T_{AB}^2(t;\rho_0)\big],
\end{equation}
plotted in Fig.~\ref{fig:Timelike_EE}(b). We find that the enhanced extraction windows of TQET are synchronized with the build-up of these timelike correlation diagnostics: the times $t_{\min}$ at which the optimized net advantage $\Delta E_{\min}(t)$ attains local minima (see Fig.~\ref{fig:DQET_energy}(d)) coincide, within numerical resolution, with the critical points (extrema) of $\Tr[T_{AB}^2(t;\rho_A)]$ and of $\Delta \Tr T^2(t)$. Figure~\ref{fig:Timelike_EE}(c) quantifies this synchronization by comparing $t_{\min}$ against the corresponding critical times extracted from the correlation diagnostic, showing that the points cluster near the diagonal.

Taken together with the analytical optimization condition for TQET, these observations support the following physical picture: once time-separated correlations between $A$ and $B$ become appreciable, the ``predictor'' term governing the optimal conditional drive at $B$ becomes large, opening time windows where conditional feedback yields a substantial extraction advantage over natural time evolution.

\section{\label{sec:conclu}Conclusion and discussion}
In our study, we introduce a novel protocol called Timelike Quantum Energy Teleportation (TQET) designed to transport quantum energy across spacetime. We have rigorously proven that the amount of energy gained through TQET is always greater than or equal to that obtained via natural time evolution for any spin chain where $[H_A, H_B] = 0$. TQET exploits both the temporal and spatial quantum correlations between agents separated by time and space, considerably enhancing the efficiency of energy transport beyond existing methods such as Quantum Energy Teleportation (QET) and natural time evolution. In our demonstration using the Ising model, TQET increases energy efficiency from about 3\% to approximately 40\%, marking a more than 13-fold improvement over QET. Furthermore, we validated certain benefits over natural time evolution and confirmed that the maximum relative energy induced by TQET, when adjusted for natural time evolution effects, is 11.6 times greater than that achieved by QET. These findings will serve as a benchmark for future research.

Further research could aim to refine the TQET protocol to boost its energy transport efficiency beyond current levels. This may involve exploring alternative quantum states or configurations to enhance temporal and spatial correlations. Additionally, applying the protocol to various observables, beyond just energy, would be fascinating. This approach suggests that any observable could be activated through quantum feedback control \cite{10.1093/ptep/ptae192}.

It will be also interesting to explore integrating TQET with other developing quantum technologies, such as quantum communication networks or quantum cryptography \cite{Ikeda:2023yhm,2023arXiv230111884I}, to build comprehensive quantum information systems that harness the full potential of quantum energy teleportation.

To optimize the design of TQET and to achieve optimal resource allocations, investigating a dynamical multi-agent model is anticipated to be highly advantageous. Within this framework, a multitude of energy suppliers and recipients engage in strategic interactions reminiscent of game-theoretical models \cite{ikeda2025quantum}. Such an approach will elucidate the mechanisms of energy resource allocation and coordination among diverse agents, thereby maximizing overall system efficiency. Through the rigorous analysis of agent dynamics and strategic decision-making processes within this multi-agent context, novel methodologies may be developed to enhance the performance and resilience of the TQET protocol. Such advancements have the potential to catalyze useful applications in sectors requiring optimal energy distribution and utilization via quantum media.

Further theoretical exploration could delve into the fundamental principles underlying TQET. This may reveal new insights into temporal entanglement and energy activation mechanisms, which could have applications in other areas of quantum research. The Ising model, a known platform for studying quantum scrambling and chaos \cite{PhysRevLett.106.050405,PhysRevA.97.042330,PhysRevLett.121.264101,PhysRevD.104.074518,PRXQuantum.5.010201}, presents a compelling angle from which to examine TQET. Although the current results (see Fig.~\ref{fig:DQET_energy} (c)) do not make this apparent, investigating TQET from this perspective would be intriguing.

Finally, while we investigated a relation between TQET and temporal correlations, using the \textit{timelike density matrix} developed in \cite{Milekhin:2025ycm}, it would also be worthwhile to explore different measures, including timelike entanglement entropy \cite{Doi:2023zaf,PhysRevLett.130.031601} and temporal entanglement entropy \cite{Grieninger:2023knz,Bou-Comas:2024pxf,Carignano:2024jxb}.

\section*{Code availability}
A code utilized for the demonstration is accessible on GitHub \cite{Ikeda_Quantum_Energy_Teleportation_2023}.
\section*{Acknowledgments}
The author thanks Shlomi Dolev, Masahiro Hotta, Adam Lowe, Yaron Oz for fruitful collaborations and discussions on quantum energy teleportation. This work was partially supported by the NSF under Grant No. OSI-2328774. 

\bibliographystyle{utphys}
\bibliography{ref}
\end{document}